\documentclass[conference]{IEEEtran}

\usepackage{graphicx}
\usepackage{url} 
\usepackage{authblk}

\usepackage{pifont}
\newcommand{\cmark}{\ding{51}}%
\newcommand{\xmark}{\ding{55}}%

\begin{document}

\title{CrowdAL: Towards a Blockchain-empowered Active Learning System in Crowd Data Labeling}

\author[1,2]{Shaojie Hou}
\affil[1]{Multiscale Networked Systems, University of Amsterdam, The Netherlands}
\affil[2]{Vrije Universiteit Amsterdam, Amsterdam, The Netherlands}

\author[1]{Yuandou Wang}

\author[1]{Zhiming Zhao}

\maketitle 

\begin{abstract}
Active Learning (AL) is a machine learning technique where the model selectively queries the most informative data points for labeling by human experts. Integrating AL with crowdsourcing leverages crowd diversity to enhance data labeling but introduces challenges in consensus and privacy. This poster presents CrowdAL, a blockchain-empowered crowd AL system designed to address these challenges. CrowdAL integrates blockchain for transparency and a tamper-proof incentive mechanism, using smart contracts to evaluate crowd workers' performance and aggregate labeling results, and employs zero-knowledge proofs to protect worker privacy.
\end{abstract}


\IEEEpeerreviewmaketitle

\section{Introduction}

Nowadays, data is often described as the gold and blood that fuels the global economy and powers innovation across all sectors. While modern society continues to generate increasing volumes of data, the majority remains raw and unlabeled, presenting challenges and opportunities for businesses and researchers alike. Effective quality control in data labeling is crucial for ensuring the accuracy and reliability of advanced data analytics. 

Active Learning (AL) is a data-efficient method combining machine learning with expert knowledge for data labeling. An active learner may begin with a small number of unlabeled data and the selected data points are presented to an oracle (e.g., a human expert) who provides the true labels for these samples to train an initial model on these labeled data~\cite{mozafari_scaling_2014}. Employing crowd workers in AL has harnessed the benefits. These approaches involve querying labels from multiple low-cost annotators with various expertise through crowdsourcing platforms instead of relying on a high-quality oracle. 

However, there are still several challenges regarding the effects of annotation noise caused by imperfect annotators. For example, how to achieve an annotation consensus with crowd diversity? How can we protect the crowd worker's identity while ensuring transparency and trust in the crowd data labeling process? Although blockchain has been used in many crowdsourcing scenarios, integrating blockchain and AL for crowd data labeling arises new technical challenges. 

This poster introduces a novel approach called CrowdAL, which utilizes a blockchain framework and Zero Knowledge Proofs (ZKPs) to cope with these challenges in crowd AL for data labeling. We design smart contracts to provide a tamper-proof means to handle user activities publicly, ensuring data integrity and distributed consensus. Blockchain's inherent support for cryptographic functionalities also facilitates worker privacy. The results have showcased the feasibility of our approach.

\section{Related Work}

While current research has explored using blockchain to enhance trust and transparency in crowdsourcing-based label aggregation and reward distribution, it still needs to address the integration with active learning processes fully. We also notice a gap in balancing privacy and publicity. A comparison of current research integrating crowdsource labeling and blockchain is shown in Table \ref{table:related_work_comparison}. This comparison evaluates whether the proposed systems support active learning, can be deployed on a public blockchain allowing public participation, and include a fair and transparent incentivization mechanism.

\begin{table}[!htb]
\centering
\caption{Comparison of different blockchain-backed frameworks.}
\begin{tabular}{|l|c|c|c|c|}
\hline
\textbf{Paper} & \textbf{Active Learning} & \textbf{Public} & \textbf{Fairness}  \\
\hline
TFCrowd\cite{li_tfcrowd_2021}       & \xmark & \xmark  & \cmark    \\
\hline
CrowdBC\cite{li_crowdbc_2019}       & \xmark  & \cmark & \xmark  \\
\hline
ZebraLancer\cite{lu_zebralancer_2018}   & \xmark & \cmark  & \cmark \\
\hline
BFC\cite{zhang_blockchain-based_2019}           & \xmark &\cmark  & \cmark   \\
\hline
This Work     & \cmark & \cmark & \cmark \\
\hline
\end{tabular}

\label{table:related_work_comparison}
\end{table}

\section{The CrowdAL Framework}

\subsection{Architecture}
\begin{figure}[!htb]
    \centering
    \includegraphics[width=\linewidth]{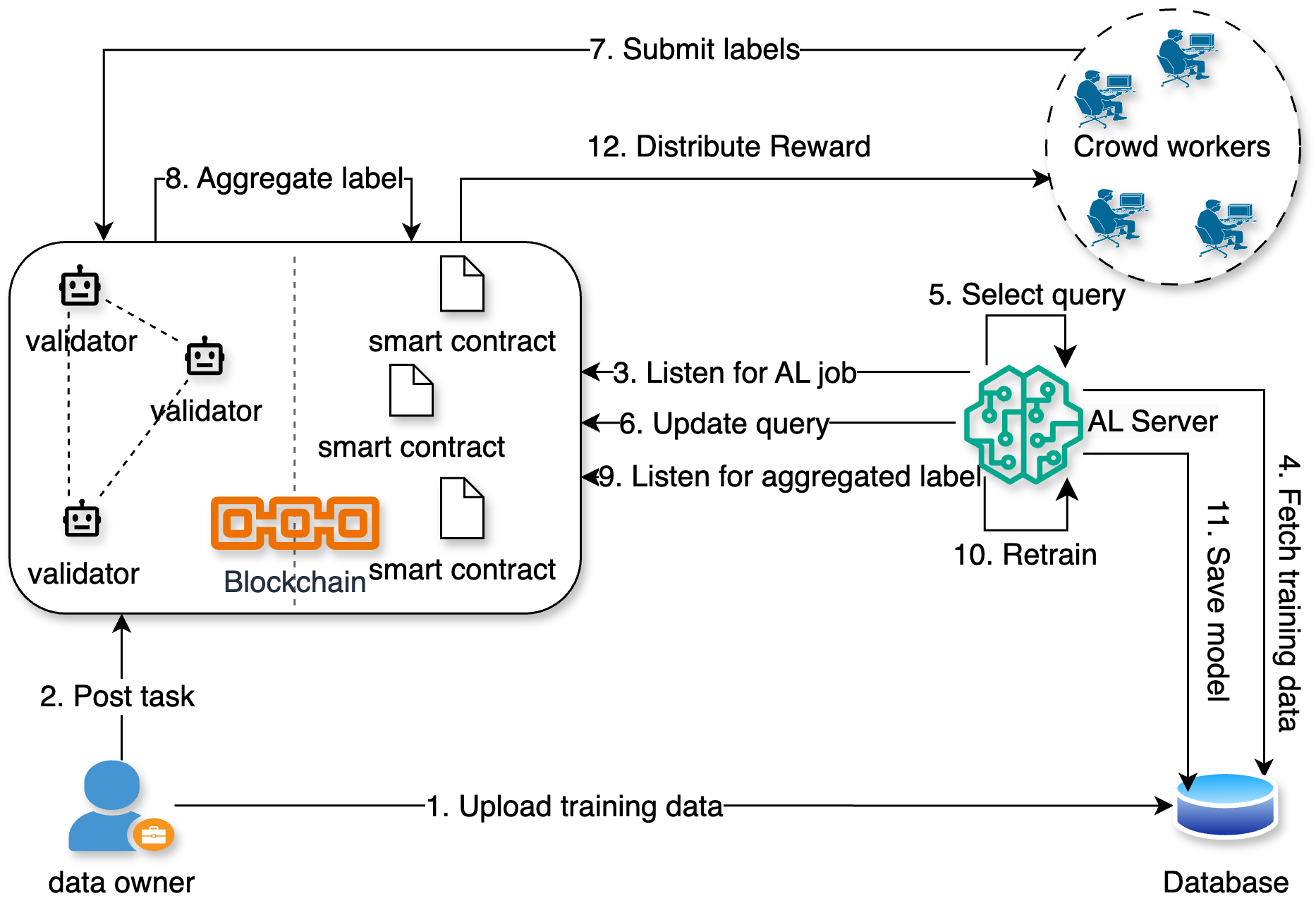}
    \caption{An overview of the CrowdAL framework and the interaction workflow.}
    \label{fig:interaction}
\end{figure}

We propose CrowdAL, a decentralized platform for data labeling in AL and the source code is available at GitHub\footnote{\url{https://github.com/ShaojieH/zktree-al/}}. We design the system architecture and interactions between the system components are depicted in Fig.~\ref{fig:interaction}. 
CrowdAL contains two components: smart contract factory and AL server. Contract factory is responsible for contract template storage, contract generation and contract deployment, while AL server is responsible for managing AL process, e.g., label querying and model training. The workflow begins with a user interacts with JobManagement contract. AL server listens for job creation event, retrieves unlabeled data from the database, and starts the training process. Users then submit labels through JobInstance contract, while AL server listens for aggregated labels. Once training completes, AL server updates the model and stores it in the database. Finally, rewards are distributed based on contribution of workers.

Note that label aggregation and incentivization in CrowdAL are executed through smart contracts to ensure fairness and transparency. \texttt{JobInstance} contract collects labels submitted by workers and uses majority voting to aggregate these labels into a single result. When distributing rewards, contribution or performance of each worker is evaluated by comparing their submitted labels to truth value, which are predicted by the trained model. Based on this evaluation, rewards are distributed proportionally to the accuracy of the workers' contributions.

We introduce ZKPs to maintain privacy in our system. Via employing the Groth-16~\cite{groth_size_2016} ZKP, it can achieve (1) anonymous labeling where the label submitted can not be associated with the identity of the worker and (2) anonymous performance evaluation where a worker's performance can be counted without revealing the labels he submitted. Specifically, we use a commit-nullify scheme to prove membership and disassociate workers from their labels. When submitting labels, workers send commitments when they join the job and prove they know the pre-image of a commitment. Since a worker can send his labels along with the proof from any address, it is said that he submits labels anonymously. Besides, when submitting each label, the worker nullifies the join job commitment and sends a single-use label commitment. In the performance evaluation phase, these label commitments are used to generate a performance proof, where a worker could prove his ownership of the labels he submitted while evaluating his performance by comparing his labels to the publicly available truth.



\subsection{Experiments and Results}
\begin{figure}[!htb]
    \centering
    \includegraphics[width=\linewidth]{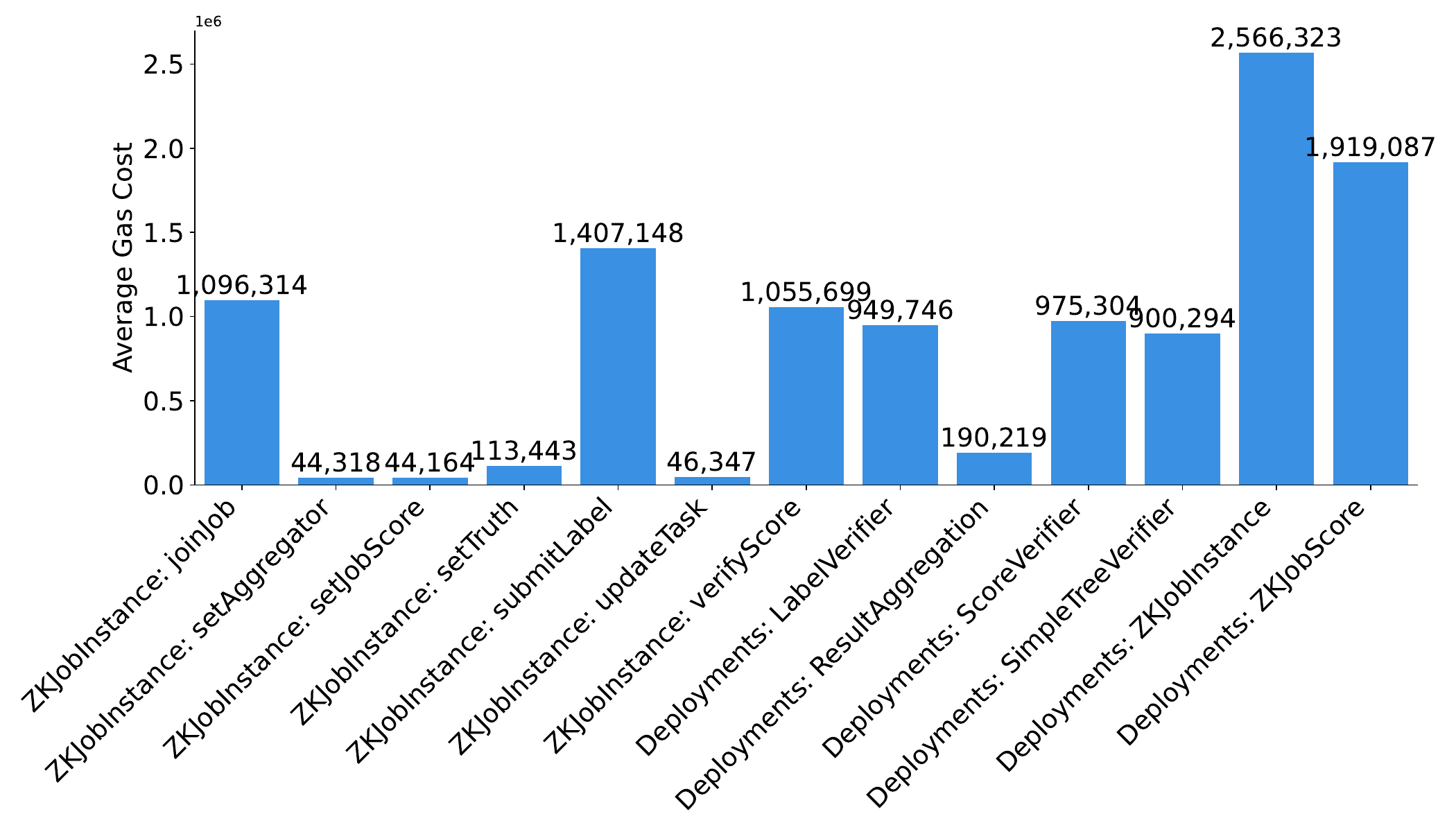}
    \caption{Comparisons of gas usage among smart contract deployment. }
    \label{fig:gas}
\end{figure}
We implemented the CrowdAL smart contracts in solidity 0.8.24 and deployed them on a local Ethereum chain to validate our approach and measured gas usage for smart contract function calls. As shown in Fig.~\ref{fig:gas}, six smart contracts have been successfully deployed and generate gas consumption. ZKP-related contract deployment and related functions, such as \texttt{ZKJonInstance} and \texttt{ZKJobScore} are more gas-intensive than others. Hence, optimizing those functions could potentially lower gas usage in the smart contract deployment and execution.

\section{Conclusion}

This poster presents CrowdAL, a blockchain-empowered crowd active learning system that addresses the challenges of integrating AL with crowd data labeling. This work has explored the implementation of a decentralized active learning system utilizing blockchain technology, while balancing transparency and privacy via zero-knowledge proofs.


\section*{Acknowledgment}
This work is part of a Master's thesis from the VU\&UvA Master thesis program. Also, this research was made possible through partial funding from several European Union projects: CLARIFY (860627), ENVRI-Hub Next (101131141), EVERSE (101129744), BlueCloud-2026 (101094227), OSCARS (101129751), LifeWatch ERIC, BioDT (101057437, through LifeWatch ERIC), and Dutch NWO LTER-LIFE project. 

\bibliographystyle{unsrt}
\bibliography{references}

\end{document}